# Factors Influencing Rural Female Workforce Participation in India, 2017-2025: An Econometric Analysis Using District-Level Panel Data

Bishwanath Goldar

Non-Resident Senior Fellow, Institute of Economic Growth, Delhi

Email: bgoldar49@iegindia.org, b_goldar77@yahoo.com

Suresh Chand Aggarwal

Visiting Professor, Institute for Human Development, New Delhi

Email: sureshchag@yahoo.com

**Abstract**: Rural female workforce participation rate (RFWPR) in India declined between 1993-94 and 2017-18. After 2017-18, the trend reversed, and RFWPR rose from 23.7% in 2017-18 to 44.9% in 2025. Using district-level panel data for about 500 districts for nine years, 2017-18 to 2025, the paper analyzes econometrically some key determinants of RFWPR in India in the recent period. We find that the presence of young children in the family tends to reduce RFWPR, consistent with findings from several previous studies. Our econometric model results indicate that an increase in household real income, proxied by the monthly per capita consumption expenditure deflated by the consumer price index, tends to reduce RFWPR. Increases in the real wage rate of casual workers in rural areas, likely reflecting a shortage of such labour, particularly in agriculture, tend to raise RFWPR. However, the positive effect of wage hikes on RFWPR continues only until real rural wages reach a particular level. Beyond that wage rate, the positive effect of wages on RFWPR diminishes or could even turn negative, i.e., rural wage hikes tending to reduce RFWPR.

Keywords: Rural female workforce participation; India's rural economy; Impact of household income; Impact of wage hikes; District-level panel data

JEL codes: J21; J22

Corresponding author: Bishwanath Goldar, Non-Resident Senior Fellow, Institute of Economic Growth, Delhi, India. Email: bgoldar49@iegindia.org, b_goldar77@yahoo.com

# Factors Influencing Rural Female Workforce Participation in India, 2017-2025: An Econometric Analysis using District-Level Panel Data

## 1. Introduction

Female workforce participation rate (FWPR) for ages 15 and above fell from 32% in 1993-94 to about 20% in 2017-18 (estimates based on the concept of usual principal status) (Goldar and Aggarwal, 2026a). In studies dealing with trends in employment in India, this fall in FWPR in the period after 1993-94 has been attributed to several factors including rising educational enrollment of females (Chadha, 2003; Kapsos et al., 2014; Mehrotra et al., 2014; Thomas, 2015; Ghose, 2016), rising household income and decline in poverty (Kapsos et al., 2014; Ghose, 2016; Mehrotra and Sinha, 2017), agricultural mechanization, and heavy unpaid domestic and care work responsibility on females (Thomas, 2015), and lack of suitable non-farm jobs, coupled with a shift towards the use of capital-intensive technology in such industries (Kapoor, 2016; Das et al., 2015).

Since 2017-18, FWPR in India has trended in the opposite direction, a reversal widely acknowledged among researchers studying employment in India. PLFS (Periodic Labour Force Survey, NSSO, Ministry of Statistics and Programme Implementation, Government of India) data show a significant increase in FWPR, particularly in rural FWPR (RFWPR), after 2017-18. Female workforce participation rate in the usual principal and subsidiary status for persons aged 15 years and above rose from 22% in 2017-18 to 40% in 2025. RFWPR rose from 23.7% in 2017-18 to 44.9% in 2025. The increase in RFWPR has been accompanied by a large increase in the number of unpaid female family workers (also called helpers) in rural areas.[1] However, the female workforce participation rate in rural areas rose post-2017, even when unpaid work is excluded (Ravi and Kapoor, 2024).

[1] Out of the incremental employment of rural females between 2017-18 and 2025, unpaid family workers form a large part. However, the share of unpaid family workers among rural females declined from 38.6% in 2017-18 to 32.8% in 2025. The share of own-account workers among rural female workers increased from 18.6% in 2017-18 to 37.2% in 2025 (Goyal et al., 2026). Considering all self-employed workers, which include unpaid family workers and own-account workers, their share of total rural female employment rose from about 57% to about 70%.

The factors that were put forward by various researchers to explain the decline in FWPR between 1993-94 and 2017-18 do not explain the recent significant surge in RFWPR. For instance, increases in female educational enrollment were considered to be a factor behind the fall in FWPR between 1993 and 2017. However, female educational enrollment has not declined post-2017, which, if true, could explain recent hikes in RFWPR. Instead, the Gender Parity Index, which shows the ratio of female-to-male students, has improved in recent years. Similarly, if agricultural mechanization was one of the reasons for the decline in RFWPR between 1993 and 2017, as argued by Thomas (2015), who considered the period up to 2011-12, the situation in rural areas in this regard has not changed drastically in the recent period (see, for example, Mehta et al., 2023) to warrant a reversal of the trend in RFWPR.

Some recent studies have claimed that the observed hikes in RFWPR after 2017-18 are principally driven by economic distress in rural areas (Singh and Ankita, 2025; Thakur and Chaudhary, 2025; Baruah and Ankita, 2026; Ghatak et al., 2026).[2] Other studies have sought an explanation of rising RFWPR in other factors or developments in India's rural economy (see, for example, Goldar and Aggarwal, 2024, who draw attention to the growing shortage of labour in rural areas and increasing shifting of male workers from agriculture to non-agricultural activities, drawing attention also to improvements in domestic piped water supply that reduced, to some extent, the rural females' burden of domestic work, enabling them to find more time to take up productive work).[3] Sinha (2025) claims that the skilling support (farm and non-farm) and credit access for diversified livelihoods to women's collectives under the *Deendayal Antyodaya Yojana National Rural Livelihood Mission* (DAY-NRLM) and rising female employment in the Mahatma

[2] Previously, the same explanation for the rural female employment surge, tracing it to rural economic distress, was provided in the *State of Working India Report, 2023 (State of Working India, 2023, Social Identites and Labour Market Outcomes,* Centre for Sustainable Development, Azim Premji University, Bengaluru, Karnataka). Ara and Shrivastav (2025) have studied employment trends up to 2022-23. They conclude that the recent increase in the LFPR (*labour force participation rate*) of women in India is driven by the distress situation induced by the COVID-19 pandemic, rather than favourable growth.

[3] Also see Goldar and Aggarwal (2026b, c), who discuss whether productivity of rural female workers is relatively low (connected with the distress-driven rural female employment hike hypothesis) and, more generally, the validity of the economic-distress-driven surge in rural female employment in India in the post-2017 period. A related issue is gender discrimination in wages and earnings in India's rural areas, which should be evaluated in the context of the recent surge in female employment in rural India, since reduced wage/earnings discrimination could have contributed to increased female workforce participation. See, in this context, Mitra and Sharma (2026).

Gandhi National Rural Employment Guarantee Scheme (MGNREGS) are the real explanation for the improvement observed in female labour force participation in India in recent years. *Economic Survey 2023–24* (Ministry of Finance, Government of India, New Delhi) attributed the increase in rural women's employment to a number of factors, including continuous high growth in agricultural output and freeing up of women's time for productive work due to substantial expansion of access to basic amenities such as piped drinking water, clean cooking fuel, sanitation, etc.

This paper uses econometric analysis to understand the role that rural household characteristics and rural labour market conditions played in increasing RFWPR in the post-2017 period. Several econometric studies have examined India's rural women's decision to participate in the labour force and workforce using recent data (for example, Gupta, 2024; Ravi and Kapoor, 2024; Goldar et al., 2025).[4] The present study differs from them. This study uses a district-level panel dataset (2017-18 to 2025) and employs panel data econometric models to examine key determinants of RFWPR.

The paper is organized as follows. The next section discusses the data, variables and econometric models. Section 3 presents the econometric results. Finally, Section 4 summarizes and concludes.

## 2. Data and Methodology

### *2.1 Data and Variables*

The analysis is based on a district-level panel dataset we constructed for rural areas across different districts in India. Unit-level data from PLFS for 2017-18 through 2023-24 are used, along with PLFS for 2024 and 2025 (the latter two are for calendar years). Five variables are considered for the analysis: (a) rural female workforce participation rate (RFWPR),[5] (b) monthly per capita consumption expenditure which is expressed at 2022-23 prices after deflating it by the CPI

[4] Other studies include Sorsa et al. (2015) and Sarkar et al. (2019). These studies considered the period before the post-2017 surge in FWPR.

[5] The workforce participation rates have been computed for persons of all ages, not confined to ages 15 and above. The same applies to the computed unemployment rate. In our opinion, this makes no major difference to our econometric findings.

(consumer price index, rural)(RMPCE) , (c) unemployment rate (UR) among females in rural areas, (d) the number of children of age five years or less in a household as a ratio to the total number of household members (CHILDSHARE), and (e) the average wage rate of male casual workers in the rural areas of the district deflated by the CPI (consumer price index, rural) and expressed at 2022-23 prices (RWAGE).[6] Some district-year observations have very low estimates of the average wage rate of casual workers (compared to the general trend). Hence, we winsorized this variable at the 1st percentile.

Using unit-level PLFS data, we computed the average values of the five variables mentioned above for each district, using the sample multipliers (provided in the unit-level PLFS datasets) as weights. These values are thus the district-level estimates of the variables used for subsequent econometric analysis.[7]

The dataset constructed for the analysis has over 630 districts. The number of districts increases over time, from 632 in 2017-18 to 732 in 2025. This increase in the number of districts is mostly due to bifurcation and/or reorganization of districts. To address this issue, we conducted our analysis on a balanced panel in which data on all five variables are available for all districts for the study period 2017-18 to 2025.[8] We have removed the districts belonging to three states, Meghalaya, Sikkim and Telangana, since there have been substantial changes in the organization of districts in these states. Thus, after applying the above criteria, the balanced panel includes 509 districts.[9] The summary statistics for the five variables are shown in Table 1.

---

[6] Observations on the average wage rate of casual female workers in rural areas are considerably fewer in the dataset compared to observations on the wage rate of casual male workers. For this reason, the analysis uses the wage rate of male casual workers instead of that of female casual workers. The wage rates of female casual workers and male casual workers are significantly correlated (see Annexure-A). Therefore, the wage rate of male casual workers is treated as a proxy.

[7] There is an issue regarding the reliability of district-level estimates of variables because the survey design of PLFS is not at the district level, except for PLFS for 2025. This point is discussed later in the paper.

[8] The issue is not fully resolved by taking a common set of districts. We discuss this point further later in the paper.

[9] The data on casual workers' wage rates could not be obtained for some observations (some districts in some years). This is another reason our balanced panel includes data for 509 districts, while a balanced panel dataset on RWFPR can be prepared for 568 districts.

**Table 1: Summary Statistics, variables used for the analysis (district-wise panel for 9 years)**

| Variable | No. of observations | Mean | Standard deviation | Minimum | Maximum |
|---|---|---|---|---|---|
| Rural female workforce participation rate (RFWPR) (%) | 4581 | 28.9 | 15.9 | 0 | 86.6 |
| Real MPCE, Rs at 2022-23 prices (RMPCE) | 4581 | 2031 | 822 | 283 | 6902 |
| Unemployment rate among rural females (%) (UR)_ | 4581 | 0.87 | 1.6 | 0 | 20.2 |
| Children up to 5 years of age divided by total number of household members (%) (CHILDSHARE) | 4581 | 2.92 | 1.54 | 0 | 19.43 |
| Average wage rate of male casual workers in rural areas, CPI deflated, Rs/day at 2022-23 prices (RWAGE) | 4581 | 382 | 125 | 141 | 1059 |

Source: Authors' computations based on PLFS data.

*District-wise distribution of RFWPR*

Next, we consider the distribution of RFWPR across districts in our data. Figure 1 compares the kernel density distribution of RFWPR among three years, 2017-18, 2019-20, and 2025. Figure 1 shows that the 2017-18 distribution is highly skewed, becomes less skewed over time, and approaches a normal distribution by 2025. The measure of skewness ($\mu_3$) is 0.67 for 2017-18, 0.25 for 2019-20 and -0.03 for 2025.

The Shapiro-Wilk W-test for normality shows that the distributions of RFWPR observations for 2017-18 and 2019-20 are not normally distributed. The 2025 distribution also does not pass the normality test. The 2024 distribution does not pass the test when all 509 districts are considered. But the distribution for 2024 is found normal when a subset of districts is chosen based on the criterion that the district-level estimates are based on at least 50 surveyed households (see Figure 2) – a point discussed later in the paper.

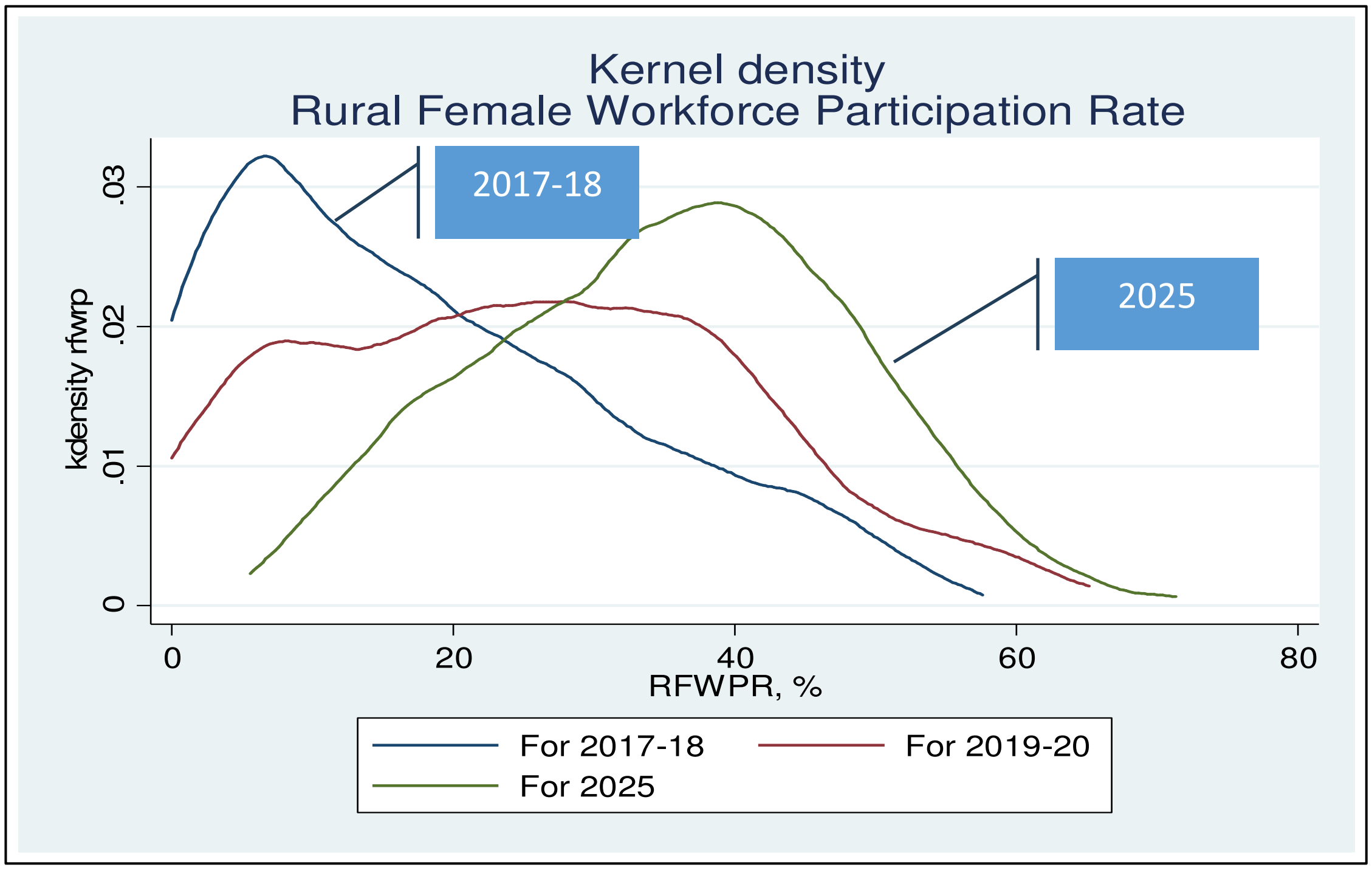

**Figure 1: Kernel Density, RFWPR, 2017-18, 2019-20 and 2025**

Source: Authors' computations based on PLFS data.

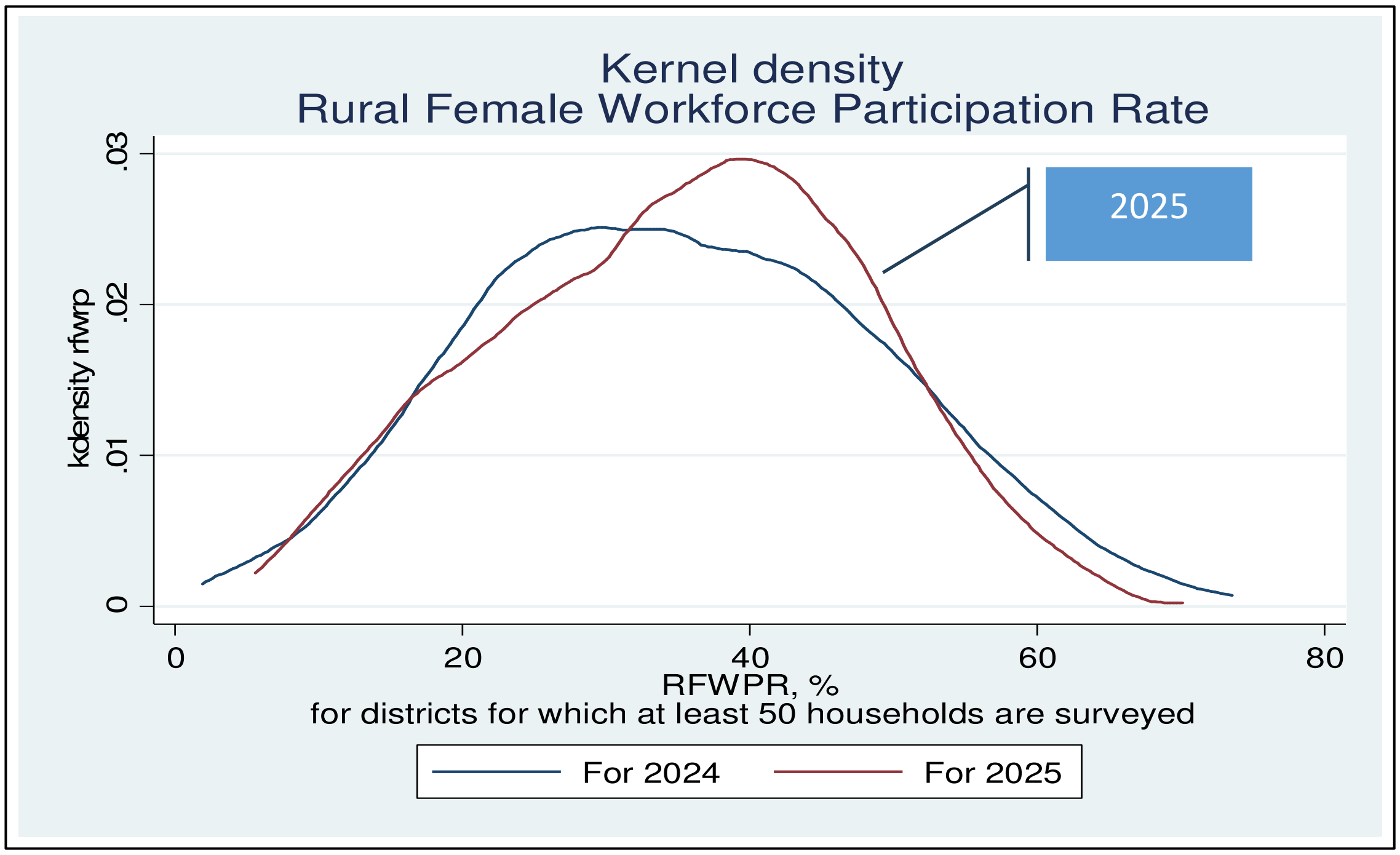

**Figure 2: Kernel Density, RFWPR, 2024 and 2025, selected districts**

Source: Authors' computations based on PLFS data.

*Panel Unit-Root Tests*

As stated above, our main variable of interest is RFWPR. We consider four explanatory variables: RMPCE and CHILDSHARE, which represent household characteristics, and UR and RWAGE, which capture labour market conditions in the district's rural areas. As part of our econometric analysis, we consider the long-term (co-integrated) relationship between these variables. Therefore, we conduct unit-root tests for the five key variables. Annexure B is devoted to panel unit-root tests. The results suggest that UR and CHILDSHARE are I(0), and RFWPR, RMPCE and RWAGE are I(1).

*Trends in the five variables*

Table 2 shows the mean values of the previously mentioned five variables for different years in the balanced panel. The table helps judge the trends. The table shows an upward trend in RFWPR, RMPCE, and RWAGE. Real per capita consumption expenditure increased by about 25% between 2017-18 and 2025. Real wages for male casual workers in rural areas increased by about 20% between 2017-18 and 2025, likely reflecting a growing shortage of such labour (Goldar and Aggarwal, 2024). For the other two variables, unemployment rate among rural females and the ratio of young children to total household members, there is no upward or downward trend.

The estimates of RMPCE should be interpreted with caution. These estimates of household consumption used in the PLFS are for classification purposes and are not as robust as estimates based on household consumption expenditure surveys. However, the average RMPCE estimates for different years in Table 2 are consistent with study findings from consumption expenditure surveys on the observed trends. Several studies have noted a significant decline in poverty in India between 2011-12 and 2022-23/2023-24 (Bhalla et al., 2023; Rangarajan and Dev, 2024; Bhalla and Bhasin, 2024).[10] Our estimates in Table 2 point in the same direction.

[10] A recent World Bank study (*Poverty and Equity Brief*, October 2026; authors: A. Wambile, D. Lestani and N. Kochhar) reports that between 2011 and 2023, the share of the population in India living below the Lower Middle-Income Country poverty line ($4.20/day, 2021 PPP) fell from 57.7% to 16.1%. The share of the population facing extreme poverty ($3.00/day) fell from 27.1% to 2.6%. The trends observed in average real MPCE in rural areas are consistent with the trends in poverty in India noted by several studies.

**Table 2: Year-wise mean of five key variables, balanced panel dataset**

| Year | RFWPR ((%) | RMPCE (Rs at 2022-23 prices) | CHILD SHARE (%) | UR (%) | RWAGE (Rs per day at 2022-23 prices) |
|---|---|---|---|---|---|
| 2017-18 | 18.1 | 2084.0 | 2.98 | 0.87 | 338.8 |
| 2018-19 | 20.0 | 2141.6 | 3.04 | 0.92 | 363.1 |
| 2019-20 | 25.4 | 2186.6 | 2.95 | 0.92 | 367.4 |
| 2020-21 | 28.3 | 1608.0 | 2.82 | 0.78 | 371.0 |
| 2021-22 | 27.9 | 1692.3 | 2.96 | 0.76 | 402.8 |
| 2022-23 | 33.3 | 1749.3 | 2.98 | 0.79 | 398.4 |
| 2023-24 | 36.1 | 1820.2 | 2.86 | 0.98 | 395.4 |
| 2024 | 35.4 | 2461.5 | 2.87 | 0.93 | 388.8 |
| 2025 | 35.4 | 2538.2 | 2.83 | 0.87 | 414.0 |

Source: Authors' computations based on PLFS data.

*2.2 Econometric model*

The econometric model is specified as follows (the term $a_i$ allows for district effects and $\lambda_t$ allows for year effects):

$$\Delta RFWPR_{it} = a_i + \lambda_t + \lambda RFWPR_{i,t-1} + \alpha\, CHILDSHARE_{it} + \beta UR_{i,t-1} + \sum_{r=1}^{2} \gamma^r \Delta\, RWAGE_{i,t-r} + \sum_{s=1}^{3} \theta^s \Delta RMPCE_{i,t-s} + \xi_{it} \ldots (1)$$

The dependent variable is the change in RFWPR between two successive years. If the level of female workforce participation reached is already relatively high, further increase may be difficult. This is why we include one-year lagged RFWPR as an explanatory variable. The coefficient of this variable is expected to be negative.

We use the share of children among household members (current year) as an explanatory variable. The coefficient of this variable is expected to be negative (because of the burden of care work).

We use the unemployment rate among rural females in the previous year as another explanatory variable. This variable is lagged by one year to address potential endogeneity. If the unemployment rate was high in the previous year, many rural women were willing to work but could not find a suitable job; hence, this is likely to raise the rate of increase in RFWPR. Thus, we expect a positive coefficient.

For the real wage variable, we use a difference form since it is integrated of order one, i.e., I(1), whereas the dependent variable (ΔRFWPR) is I(0). The changes in the real wage rate (RWAGE) in t-1 and t-2 are taken as factors influencing ΔRFWPR in year t. The model excludes the current-year wage increase because of potential endogeneity. An increase in real wages may signal a causal labour shortage in rural areas of the district. Hence, a higher value of ΔRWAGE in years t-1 and t-2 is expected to lead to a bigger increase in RFWPR in year t.

Since RMPCE is I(1) in our assessment, this variable is taken in difference form. In this case, endogeneity is again an issue, so we cannot use the current-year value. Accordingly, we use RMPCE for t-1, t-2, and t-3.

It is hard to say a priori what the sign of the coefficients on the MPCE variable (an indirect measure of household income) will be. Goldar et al. (2025) find that adult women in rural households with more than 0.2 hectares of land spend more time on productive work than those in households with less than 0.2 hectares of land or no land. This suggests a positive effect of household land assets (and hence household income) on the likelihood of women in a household participating in the workforce. On the other hand, it is commonly believed that increased income in rural households beyond the subsistence level is likely to lead rural women to withdraw from the labour force and workforce (which was offered as the explanation for the fall in female labour force and workforce participation in the 1990s and the next two decades). Studies have found a negative effect of income on female participation in the labour force and workforce in India (Sorsa et al., 2015; Sarkar et al., 2019). Evidently, the impact of MPCE on RFWPR is hard to envisage and is an empirical question.

The equation specified above has been estimated by the system GMM (Generalized Method of Moments) estimator. While using lagged values of some explanatory variables addresses, to some extent, the problem of endogeneity, the GMM estimator is a better solution to potential endogeneity. This is why the GMM estimator is preferred.[11]

To supplement the analysis based on equation (1) above, a second econometric exercise has been done in which the dynamic fixed effects (DFE) model is estimated, which provides a long-run (co-integrated) relationship among the five variables, along with the relationship among them in the short run.[12] In this case, the variables in level form are used in the long-term relationship. However, endogeneity remains an issue. Hence, RMPCE, the real wage rate and the unemployment rate are taken with a one-year lag in the long-run equation. i.e., $RFWPR_t$ is taken as a function of $RMPCE_{t-1}$, $RWAGE_t$, $UR_{t-1}$ and $CHILDSHARE_t$.

*2.3 District-level variable estimates – reliability and inter-temporal comparability*

A question can be raised about the reliability of the district-level estimates of the five variables used for the econometric analysis after forming a district-wise panel for nine years, on the ground that the PLFS surveys are not designed at the district level, but at the state level (except for 2025, which employed a design at the district level). Since the unit-level datasets contain district identifiers, it is possible to make district-level estimates. However, one cannot be confident in the reliability of the estimates, especially for districts with relatively few sample households.

In one set of regression results, we ignore this issue of reliability of district-level estimates of the five variables made from unit-level PLFS data. The basis for this treatment is as follows. For one year (2023-24), we have computed the simple average of each of the five variables for each district and compared it with the weighted average, which uses the sample multipliers. The correlation between the two sets of averages is very high. The correlation coefficient is 0.85 for CHILDSHARE, and it ranges from 0.96 to 0.99 for the other four variables. Thus, we could treat our dataset as a pseudo panel – each observation may be taken as the values of the variables for a representative rural household in the districts. With this interpretation made of the dataset, we can take care of the issue of reliability of district-level estimates obtained from PLFS for years prior to 2025 to a large extent, but not fully.

[11] The equation we estimate is of the form: $\Delta Y=\alpha+ \lambda Y_{t-1}+ \beta\Delta X_{t-1}+ \gamma Z + u_t$. Note that $\Delta Y=Y_t - Y_{t-1}$. Thus, the equation can be transformed into: $Y_t=\alpha+(1+\lambda) Y_{t-1}+ \beta\Delta X_{t-1} +\gamma Z +u_t$. Because the lagged dependent variable enters the regression as an explanatory variable, fixed-effects or random-effects models should not be used; instead, the GMM estimator should be used.

[12] For the short-run equation, dummy variables for the COVID-19 pandemic are used.

The second set of regression estimates better addresses the reliability of district-level estimates of the five variables of interest. In this case, we consider only those observations in which the number of surveyed households on which the district-level estimates are made is at least 50. Applying this criterion reduces the number of useful observations by about 20%. There are district-year observations that are based on data for eight households or 16 households (in about 1.5% of observations). The reliability of such district-year observations is questionable, and their use in the regression analysis may affect the results. The second dataset addresses this issue.

The application of the above criterion for district selection, i.e., considering only the districts for which at least 50 rural households were surveyed, does not cause any major change in the distribution of districts according to RFWPR. The two distributions are similar.

Another issue to which attention needs to be drawn is that of bifurcation and reorganization of districts. It has been noted above that three states have been excluded from the analysis because these underwent significant changes in district organization (i.e., boundaries of the districts). Some issues of inter-temporal comparability of the districts still remain. If a district gets bifurcated, our panel dataset contains the full district till a particular year and only a part of the district in later years. Since the variables used for the analysis are ratios (not absolute values of a parameter such as the number of persons employed), the problem of inter-temporal comparability is relatively less. However, inter-temporal comparability issues still remain. To address this issue, we ran a third set of regression estimates, restricting the analysis to states in which the number of districts did not change during the study period, 2017-2025. This reduces the sample size by another 60 percentage points; the third analysis includes about 23% of the observations in the original panel dataset used for the first set of estimates.

The three datasets discussed above are hereafter called: (a) balanced panel dataset, (b) pruned panel dataset-1, and (c) pruned panel dataset-2. As explained above, pruned panel dataset-1 is confined to those district-year observations for which we have data on at least 50 rural households,[13] and pruned panel dataset-2 is a part of pruned panel dataset-1 and contains data for those states in which the number of districts did not change during the study period, 2017-2025.

[13] The correlation between the simple average and the weighted average of variables for 2023-24 mentioned above is for the pruned panel dataset-1.

## 3. Empirical Results

### *3.1 Model Estimates Based on GMM*

Table 3 presents the estimates of equation (1) obtained by applying the GMM estimator. Three sets of estimates are presented: one based on the balanced panel (more districts and observations) and two for pruned dataset-1 and pruned dataset-2 derived from the balanced panel, as explained in Section 2.3 above.

**Table 3: Determinants of Change in RFWPR, Regression Results, System GMM Estimator, Two-step Method (Dependent variable =ΔRFWPR)**

| Explanatory variables | Balanced panel | Pruned panel dataset-1 | Pruned panel dataset-2 |
|---|---|---|---|
| RFWPR (t-1) | -0.43(0.036)*** | -0.54 (0.046)*** | -0.38 (0.053)*** |
| UR (t-1) | 0.50 (0.22)** | 0.61 (0.21)*** | 0.12 (0.41) |
| CHILDSHARE (%)(t) | -2.50 (0.75)*** | -1.88 (0.81)** | -1.66 (0.64)*** |
| ΔRWAGE (t-1) | 0.0029 (0.0039) | 0.0031 (0.0033) | 0.0067 (0.0072) |
| ΔRWAGE (t-2) | 0.0028 (0.0032) | 0.0050 (0.0030)* | 0.0030 (0.0062) |
| ΔRMPCE (t-1) | -0.0013 (0.0006)** | -0.0010(0.0006)* | -0.0023(0.0010)** |
| ΔRMPCE (t-2) | -0.0015 (0.0007)** | -0.0012 (0.0006)** | -0.0027 (0.0010)*** |
| ΔRMPCE (t-3) | -0.0015(0.0005)*** | -0.0015(0.0004)*** | -0.0021(0.0009)** |
| Constant | 23.0(2.26)*** | 24.4 (2.75)*** | 18.2(2.44)*** |
| No of observations | 2545 | 2029 | 593 |
| No. of districts | 509 | 500 | 139 |
| Wald chi-square and prob. | 414.9 (0.000) | 336.6 (0.000) | 111.9 (0.000) |
| AR(1) | 0.000 | 0.000 | 0.000 |
| AR(2) | 0.346 | 0.487 | 0.820 |
| Hansen test of over-identification, chi-square and probability | 166.7 (0.000) | 135.2 (0.018) | 130.3 (0.127) |

Source and note: Authors' computation. Robust standard errors are shown in parentheses. Variable notation given in Table 1. *, **, *** statistically significant at 10, 5 and 1 per cent levels, respectively.

The results show that the coefficient of RFWPR(t-1) is negative and statistically significant, indicating that the change in RFWPR slows when the previous year's RFWPR is already relatively high. This is as expected.

The coefficient of CHILDSHARE is negative and statistically significant. Thus, having young children (up to 5 years of age) in the family adversely affects RFWPR. This is, again, expected. Gupta (2024) finds that the mothers' participation in productive work is significantly reduced by 9.9% if young children aged 0–5 years are in the household. Our findings agree with the findings of Gupta (2024). These are also consistent with Goldar et al. (2025), who find that when children aged 10 or less are in the household, females spend more time on care work and less time on productive work.

The coefficients of the real wage variable are positive in all six cases. It is statistically significant only in one case out of six. The results suggest that an increase in real wages for casual workers tends to raise RFWPR, but the effect may not be strong.

The coefficient of RMPCE is consistently negative. In all nine cases, the coefficient is statistically significant. It appears, therefore, that the effect of an improvement in a rural household's standard of living is to reduce female workforce participation. These results align with the findings of Sarkar et al. (2019).

The constant term in the estimated model represents the impact of factors not included in the model. For instance, the freeing up of women's time for productive work due to substantial expansion of access to basic amenities such as piped drinking water, clean cooking fuel, sanitation, etc. will be reflected in the constant term. Similarly, the skilling support (farm and non-farm) and credit access for diversified livelihoods to women's collectives under DAY-NRLM will be captured by the constant term. In our results, the constant term is positive and statistically significant at the 1% level. This result provides a basis to argue that the two developments above, along with other such developments (for instance, increasing access to mobile phones for rural women in India), contributed to increases in RFWPR.

*3.2 Estimates of the Dynamic Fixed Effects Model*

Table 4 presents the estimates of the dynamic fixed effects (DFE) model. The table shows the long-run and short-run coefficients. As mentioned earlier, this model considers the long-term relationship among the variables at levels.

The error correction term ($ECM_{t-1}$) is negative and statistically significant; its value is about -0.7. Thus, if a deviation from the long-run relationship occurs, it is corrected quickly.

**Table 4: Determinants of RFWPR, Regression Results, Dynamic Fixed Effects Model (dependent variable = RFWPR)**

| Explanatory variables | Balanced panel | | Pruned panel dataset-1 | | Pruned panel dataset-2 | |
|---|---|---|---|---|---|---|
| | Long run | Short run | Long run | Short run | Long run | Short run |
| ECM (t-1), error correction | | -0.71 (0.015)*** | | -0.69 (0.016)*** | | -0.73 (0.028)*** |
| ΔUR (t-1) | | -0.076 (0.117) | | -0.076 (0.122) | | -0.025 (0.190) |
| ΔCHILDSHARE (%) | | -0.07 (0.12) | | -0.19 (0.13) | | -0.19 (0.22) |
| ΔRWAGE (t-1) | | -0.013 (0.003)*** | | -0.013 (0.003)*** | | -0.010 (0.004)** |
| ΔRMPCE (t-1) | | 0.002 (0.0004)*** | | 0.002 (0.0005)*** | | 0.0002 (0.0008) |
| Constant | | 21.6 (1.82)*** | | 19.4 (1.97)*** | | 16.8 (2.62)*** |
| UR (t-1) | 0.29 (0.23) | | 0.40 (0.25)# | | 0.16 (0.36) | |
| CHILDSHARE (%) | -0.47 (0.25)* | | -0.26 (0.31) | | -0.02 (0.47) | |
| RWAGE (t-1) | 0.030 (0.005)*** | | 0.032 (0.005)*** | | 0.022 (0.009)** | |
| RMPCE (t-1) | -0.004 (0.0006)*** | | -0.003 (0.0007)*** | | -0.0006 (0.0010) | |
| No of observations | 2545 | | 2029 | | 539 | |

Source and note: Authors' computation. Robust standard errors shown in parentheses. *, **, *** statistically significant at 10, 5 and 1 per cent levels, respectively. # prob.<0.11

The long-run coefficient of the child share variable is negative in all three regressions and is statistically significant in one of them. These results, along with the results in Table 4, indicate that the presence of young children in the household negatively affects female workforce participation.

The long-run coefficient of the real wage rate is positive in all three regressions and statistically significant in all of them. This indicates that a hike in the casual workers' wage rate in a district (presumably indicating a shortage of such labour) tends to raise RFWPR. Data presented in Table 3 above indicated an upward trend in the real wage rate of casual workers. The finding of a significant positive long-run coefficient of the real wage rate variable implies that the increase in real rural wages over time was one of the factors that contributed to hikes in RFWPR.

The long-run coefficient of real MPCE is negative and statistically significant in two cases. Similar results were obtained from the GMM estimate of equation (1). This suggests that as rural households' standards of living improve, RFWPR tends to fall in the long run. Since RMPCE rose significantly between 2017-18 and 2025 (see Table 2), it should have lowered RFWPR. However, the opposite occurred; therefore, the negative effect of increases in RMPCE may have been counterbalanced by other developments. One factor was the rise in casual workers' wages in rural areas. Also, various government schemes have contributed to the rise in RFWPR, as argued by Goldar and Aggarwal (2024) and Sinha (2024). Arguably, these developments have made a major contribution to hikes in RFWPR.

This study did not aim to test the hypothesis of an economic-distress-driven surge in RFWPR, a view expressed by several researchers. Therefore, based on the results, we cannot support or refute the said hypothesis. However, a few comments are needed, since many studies have put forward this hypothesis. In this context, it should be noted that the analysis above provides no empirical evidence that could be considered empirically supportive of the hypothesis of an economic-distress-driven surge in RFWPR. Rather, the econometric results seem to indicate the opposite.

### *3.3 Additional Analysis*

Goldar et al. (2025) found an inverted-U-shaped relationship between female wage rate and the amount of time rural females devote to productive work vis-à-vis unpaid domestic work and care work. The analysis used time-use survey data for the year 2019.[14] The threshold was found at a wage rate of Rs 170 per day, beyond which increases in the wage rate negatively affect the time rural women spend on productive work. One disadvantage of the analysis presented by Goldar et al. (2025) is that it is based on cross-sectional data for one year. Because the present study uses a panel dataset, it would be useful to see whether the same relationship holds in our dataset. The advantage of our district-level panel data is that we can control for unobserved cross-district heterogeneity.

To investigate whether the relationship between wage rate and RFWPR differs between low-wage districts and relatively higher-wage districts, we consider rural female casual workers' wage rates in 2019-20. For some districts, the wage data are not available for 2019-20, but are available for 2020-21 or 2018-19. In those cases, the missing data for 2019-20 has been filled using data for 2020-21 or 2018-19.

In the next step, we divided the districts into two groups: those in which the average wage rate of rural female casual workers was Rs 170/day or less, and the remaining districts. We also estimate the model using all districts in which the female casual workers' wage rate was above Rs 230/day, to sharpen the contrast with districts in which the wage rate was below Rs 170/day. We estimated Equation (1) using the GMM estimator for sub-samples. We consider the pruned panel datasets 1 (as explained above) for the analysis. Table 5 reports the results.

[14] Time Use in India-2019, National Statistical Office, Ministry of Statistics and Programme Implementation, Government of India.

**Table 5: Determinants of Change in RFWPR, Regression Results, System GMM estimator, two-step method (dependent variable =ΔRFWPR)**

| Explanatory variables | Pruned panel dataset-1 | | |
|---|---|---|---|
| | Districts in which female casual workers' wage rate in 2019 was Rs 170/day or less | Districts in which female casual workers' wage rate in 2019 was above Rs 170/day | Districts in which female casual workers' wage rate in 2019 was above Rs 230/day |
| RFWPR (t-1) | -0.52 (0.066)*** | -0.42 (0.057)*** | -0.30 (0.084)*** |
| UR (t-1) | 0.97 (0.61)# | 0.65 (0.39)* | 0.63 (0.48) |
| CHILDSHARE (%)(t) | -0.40 (0.92) | 0.30 (1.51) | -0.26 (1.32) |
| ΔRWAGE (t-1) | 0.0123 (0.0077)* | -0.011 (0.008) | -0.006 (0.009) |
| ΔRWAGE (t-2) | 0.0124 (0.0063)** | -0.049 (0.023)** | -0.060 (0.022)*** |
| ΔRMPCE (t-1) | 0.0015 (0.0009) | 0.0007 (0.0006) | 0.0006 (0.0008) |
| ΔRMPCE (t-2) | -0.0029 (0.0012)** | -0.0032 (0.0014)** | -0.0047 (0.0019)** |
| ΔRMPCE (t-3) | -0.0006 (0.0009) | -0.0059 (0.0015)*** | -0.0087 (0.0017)*** |
| Constant | 20.5 (3.75)*** | 12.1 (4.47)*** | 8.63 (4.09)** |
| No of observations | 541 | 1334 | 781 |
| No. of districts | 140 | 322 | 187 |
| Wald chi-square and prob. | 140.4 (0.000) | 174.4 (0.000) | 101.7 (0.000) |
| AR(1) | 0.000 | 0.000 | 0.000 |
| AR(2) | 0.760 | 0.274 | 0.364 |
| Hansen test of over-identification, chi-square and probability | 108.6 (0.36) | 46.8 (0.014) | 37.5 (0.11) |

Source and note: Authors' computations. Robust standard errors shown in parentheses. *, **, *** statistically significant at 10, 5 and 1 per cent levels, respectively. # prob <0.12

The results in Table 5 are similar to those in Table 3 and therefore do not require a detailed discussion. Table 5 shows three patterns. First, RMPCE has a negative effect on RFWPR in both low-wage districts and relatively high-wage districts,[15] but the effect appears relatively greater in the latter. Second, a high rate of unemployment among women tends to raise RFWPR in low-wage districts, but the effect is less strong in relatively higher-wage districts. Third, a significant positive effect of wage hikes on RFWPR is found for low-wage districts. But such a positive effect of wage hikes is not found for relatively higher-wage districts. Instead, the results seem to suggest that wage hikes in such districts tend to decrease RFWPR.[16] These findings on the effect of wages on RFWPR accord well with findings of Goldar et al. (2025). It appears from the results obtained that the effect of wage hikes on RFWPR is significantly positive till a wage rate of Rs 170/day at 2019-20 prices (or thereabouts) is reached, and then this marginal effect wanes beyond that wage level and may eventually turn negative.

## 4. Conclusion

After 2017-18, RFWPR in India surged, rising from 23.7% in 2017-18 to 44.9% in 2025. Using district-level panel data for about 500 districts for nine years, 2017-18 to 2025, we analyzed econometrically some key determinants of RFWPR in India in the recent period. We estimated a model explaining year-to-year changes in RFWPR across districts. We also estimated a model that provided a long-term relationship of RFWPR with the ratio of young children to household members, real per capita consumption expenditure in households, unemployment rate among rural females and the real wage rate of casual workers.

---

[15] The RMPCE(t-1) coefficient is positive for all three estimates in Table 5, but not statistically significant in any regression. Thus, the inference is drawn on the statistically significant coefficients. Note further that the sum of the coefficients of the three lagged terms of RMPCE is negative.

[16] Supporting evidence is found from two other exercises undertaken. Instead of splitting the sample by female casual workers' wages in 2019-20 (as in Table 5), we introduced a quadratic term for ΔRWAGE and estimated equation (1) for prune dataset-1. The coefficients of $\Delta RWAGE_{t-1}$ and $\Delta RWAGE_{t-2}$ are 0.13 and 0.14, both statistically significant. The coefficients of $\Delta RWAGE^2_{t-1}$ and $\Delta RWAGE^2_{t-2}$ are (-)0.00015 and (-)0.00019 respectively, both statistically significant. We have also estimated the dynamic fixed effects model after introducing a quadratic term for real wages. The long-run coefficient of the real wage rate is 0.082, and the coefficient of its squared term is (-) 0.00006, both statistically significant. These results indicate that beyond a threshold, the positive effect of real wages on RFWPR disappears.

Our analysis finds that the presence of young children in the family tends to reduce RFWPR, consistent with findings from several previous studies. Our econometric model results indicate that an increase in household real income, proxied by the monthly per capita consumption deflated by the consumer price index, tends to reduce RFWPR. Increases in the real wage rate of casual workers in rural areas, likely reflecting a shortage of such labour, particularly in agriculture, tend to raise RFWPR. However, the positive effect of wage hikes on RFWPR occurs only up to a particular level of real rural wages, not beyond.

**References:**

Ara, S. & Shrivastav, P. K. (2025), Decoding Rising Female Labour Force Participation in India: Evidence from PLFS Data, *Economic & Political Weekly*, 60 (26 &27), 59-67.

Baruah, A., & Ankita (2026). Can Agricultural Growth Cause Higher Employment among Rural Women in India? Debunking the Myth. *Economic & Political Weekly*, 61 (12), 64–71.

Bhalla, S., Bhasin, K., & Virmani, A. (2022). Pandemic, Poverty, and Inequality: Evidence from India. Working Paper no. WP/22/69, International Monetary Fund.

Bhalla, Surjit Singh and Karan Bhasin (2024). Poverty in India over the last decade. *Economic and Political Weekly*, 59(28),60-64.

Chadha, G. K. (2003). Issues in Employment and Poverty. Discussion Paper 7, Recovery and Reconstruction Department, *International Labour Office*, Geneva.

Das, D. K., Sen, K., & Das, P. C. (2015). Labour Intensity in Indian Manufacturing. In: Ramaswamy, K.V. (Ed), *Labour, Employment and Economic Growth in India*, Cambridge University Press, Delhi, pp. 153-180.

Ghatak, M., Jha, M., & Singh, J. (2026). Rise of Unpaid Family Workers: Evidence on Distress-driven Employment Growth in India. *Economic and Political Weekly*, 61 (1), 67-72.

Ghose, Ajit K. (2016), *India Employment Report 2016*, *Institute for Human Development*, Oxford University Press, New Delhi.

Goyal, T.M., Prakash, S., Gaikwad, K., & Tandon, A. (2026). The Participation Paradox: Women's Work and Time Use in India, 2017-18 to 2025. Indian Council for Research on International Economic Relations, New Delhi.

Goldar, B., & Aggarwal, S. C. (2024). Explaining the Significant Hikes in Female Workforce Participation Rate in India in Recent Years and Some Thoughts on How to Raise It Further, Working Paper No. 462, Institute of Economic Growth, Delhi.

Goldar, B., & Aggarwal, S. C. (2026a). Female Workforce Participation in India: How to Step It Up and Increase Their Employment in Non-Agricultural Activities. In: A.N. Sharma, A.K. Mehta and V. Upadhyay (editors), *Women and Work in India: Challenges, Opportunities and Perspectives for Policy*, Springer.

Goldar, B., & Aggarwal, S.C. (2026b). Is the Productivity of Unpaid Female Family Workers in Indian Agriculture Very Low?. *Economic and Political Weekly*, 61(5): 70-72.

Goldar, B., & Aggarwal, S.C. (2026c). Explaining the Post-2017 Surge in Rural Female Employment in India: Alternate Hypotheses. Available at SSRN: https://ssrn.com/abstract=7024578 or http://dx.doi.org/10.2139/ssrn.7024578.

Goldar, B., Dasgupta, P., & Aggarwal, S. C. (2025). Impact of Rural Wages on Rural Women's Decision Regarding Time Spent on Productive Work, Household Work and Care Work. *Indian Journal of Labour Economics.* 68 (4), 1317–1342.

Gupta, I. (2024). Younger children and mothers' labour supply in rural India: Evidence from fertility stopping behaviour. *Journal of Population Research*, 41:18. https://doi.org/10.1007/s12546-024-09339-w.

Kapoor, R. (2016). Technology, Jobs and Inequality: Evidence from India's Manufacturing Sector. *Working Paper No. 313, ICRIER*, New Delhi.

Kapsos, S., Bourmpoula, E., & Silberman, A. (2014). Why is female labour force participation declining so sharply in India? ILO.

Mehrotra, S. (2014). *India's Skill Challenge. Reforming Vocational education and training to harness the demographic dividend.* Oxford University Press, New Delhi.

Mehrotra, S., & Sinha, S. (2017). Explaining Falling Female Employment during a High Growth Period. *Economic and Political Weekly*, 52(39), 54-62.

Mehta, C.R., Bangale, R.A., Chandel, N.S., & Kumar, M. (2023). Farm Mechanization in India: Status and Way Forward. *Agricultural Mechanization in Asia, Africa and Latin America*, 54(2), 75-88.

Mitra, A., & Sharma, D. (2026). Women and Work: Revisiting Inequality Issues in the Informal Sector. *Economic and Political Weekly*, 61(41): 58-64.

Rangarajan, C., & Dev, S.M. (2024). Declining Poverty Ratio: A continuing trend. Hindu, May 14, 2024.

Ravi, S., & Kapoor, M. (2024). Female Labour Force Participation Rate: An Observational Analysis of the Periodic Labour Force Survey (PLFS) from 2017-18 to 2022-23. EAC-PM Working Paper Series EAC-PM/WP/34/2024, Economic Advisory Council to the Prime Minister.

Sarkar, S., Sahoo, S., & Klasen, S. (2019). Employment transitions of women in India: A panel analysis, World Development, 115, 291-309.

Singh, I., & Ankita (2025). Increasing Female Employment in India: Distress or Better Opportunities? *Economic and Political Weekly*, 60 (8), 37-45.

Sinha, A. (2025). Understanding Trends in Female Labour Force Participation Rate in Rural India: The Role of MGNREGS and DAY-NRLM (CSEP Discussion Paper 30). New Delhi: Centre for Social and Economic Progress.

Sorsa, P., Mares, J., Didier, M., Guimaraes, C., Rabate, M., Tang, G., & Tuske, A. (2015). Determinants Of the Low Female Labour Force Participation in India. Economics Department Working Papers No. 1207. OECD.

Thakur, A.N., & Chaudhary, P. (2025). Decoding the Recent Rise in Women's Work. Participation. *Indian Journal of Labour Economics.* 68, 1455–1474.

Thomas, J. J. (2015). India's labour Market during the 2000s. In: Ramaswamy, K.V. (Ed.), *Labour, Employment and Economic Growth in India,* Delhi: Cambridge University Press, pp. 21-56.

**Annexure-A: Relation between wage rates of male and female casual workers in rural areas**

There is a significant positive correlation between wage rates of male and female casual workers. We could get about 5,900 district-year observations on the average wage rate of male casual workers in rural areas and 4,600 such observations for rural female casual workers. Because data on female casual workers' wages in rural areas are unavailable for many observations, we used male casual workers' wages in the analysis presented in the paper as a proxy for female casual workers' wage rates in rural areas. The correlation coefficient between the two variables from pooled time series and cross-section data is 0.6 for the balanced panel, leaving out Meghalaya, Sikkim and Telangana for reasons given above, and confining to those observations which are based on a rural household sample of 50 or more.

To study the relationship between female and male casual workers' wage rates, we estimate a fixed-effects model, taking the logarithm of the female wage rate as the dependent variable (denoted lnFWit) and the logarithm of the male wage rate as the explanatory variable (denoted $lnMW_{it}$). The year dummy variables are included in the estimated model to capture time effects (for instance, the effect of inflation). The estimated regression equation is shown below (t-value in parentheses):

$$lnFW_{it} = \underset{(8.32)}{1.89} + \underset{(14.35)}{0.59}\, lnMW_{it} + \text{year effects} \quad R^2 \text{ (overall)} = 0.41; \text{ No. of observations} = 3{,}004$$

The coefficient of the male wage rate variable is statistically significant at the 1% level. The estimated elasticity is 0.59, which indicates that a one per cent increase in the average wage rate of male casual workers in rural areas of a district is associated with (or accompanied by) a 0.6 per cent increase in the wage rate of female casual workers in that district.

**Annexure-B: Panel Unit-Root Tests**

The results of the panel unit-root tests (unit-root tests for panel data) for the five variables considered for the study are shown in Table B.1. The panel unit root test results (LLC, IPS and Hadri) for unemployment rate and the ratio of children to household members suggest that these two variables are integrated of order zero, i.e., these two variables are I(0).

**Table B.1: Results of Unit-root tests for five variables included in the study (district-wise panel for 509 districts and 9 years)**

| Tests | Variables | | | | |
|---|---|---|---|---|---|
| | RFWPR | RMPCE | UR-RF | CHILD SHARE | REALWAGE |
| Level | | | | | |
| LLC | -40.3 (0.000)** | -18.7 (0.000)** | -55.1 (0.000)** | -44.8 (0.000)** | -29.8 (0.000)** |
| IPS | -7.4 (0.000)** | -0.59 (0.28) | -16.9 (0.000)** | -21.3 (0.000)** | -17.0 (0.000)** |
| Hadri | 47.8 (0.000)** | 16.8 (0.000)** | 0.66 (0.25)@ | -1.94 (0.97)@ | 19.0 (0.000)** |
| First difference | | | | | |
| LLC | -60.8 (0.000)** | -17.0 (0.000)** | -43.8 (0.000)** | -72.8 (0.000)** | -55.9 (0.000)** |
| IPS | -22.0 (0.000)** | -18.8 (0.000)** | -20.8 (0.000)** | -26.7 (0.000)** | -25.1 (0.000)** |
| Hadri | -9.0 (1.00) | -1.5 (0.937) | -11.7 (1.00) | -13.3 (1.00) | -11.1 (1.00) |

Notes: (a) LLC= Levin-Lin-Chu unit-root test; IPS= Im-Pesaran-Shin unit-root test; Hadri= Hadri test (in this test, the null and alternative hypotheses are reversed). (b) Probabilities shown in parentheses. ** statistically significant at the 1% level. @ A time trend is included for this test.

Source: Authors' computations based on PLFS data.

The results for RMPCE are mixed. However, one may infer that the variable is integrated of order 1, i.e., I(1), because two of the three tests (IPS and Hadri) indicate so.

For RFWPR and real wage rate, the Hadri test results conflict with the LLC and IPS tests. While the Hadri test indicates these variables are I(1), the other two tests indicate they are I(0).

Therefore, we conducted further analyses by accounting for cross-sectional dependence and running second-generation panel unit-root tests. Table B.2 presents the results.

**Table B.2: Results of Unit-root tests and cross-sectional dependence test (district-wise panel for 9 years)**

| Tests | Variables | | |
|---|---|---|---|
| | RFWPR | RMPCE | REALWAGE |
| Level | | | |
| Pesaran CIPS test | -2.17@ (critical value at 10% = -2.63) | -2.75@ (critical value at 1% = -2.98) | -2.93@ (critical value at 1% = -2.98) |
| First difference | | | |
| Pesaran CIPS test | -2.86 (critical value at 1% = -2.28) | -3.01 (critical value at 1% = -2.28) | -3.24 (critical value at 1% = -2.28) |
| Cross-sectional dependence test (statistics; probability in parentheses) | | | |
| Pesaran (2015, 2021) | 424.0 (0.000)** | 565.8 (0.000)** | 188.0 (0.000)** |
| Pesaran, Xie (2021) | -1.12 (0.262) | 0.19 (0.849) | 2.93 (0.003)** |

Notes: (a) Pesaran CIPS (Cross-sectionally augmented IPS test) is a second-generation unit-root test for panel data, based on Pesaran (2007). It allows for cross-sectional dependence. (b) The cross-sectional dependence tests are based on: (i) Pesaran, M. H. (2015). (ii) Pesaran, M. H. (2021); and (iii) Pesaran, M. H. & Xie, Y. (2021). The software module 'xtcd2' for Stata, prepared by Jan Detizen, has been used. (c) Probabilities shown in parentheses. ** statistically significant at the 1% level. @ A time trend is included for the tests done for the variable at levels.

Source: Authors' computations based on PLFS data

.

The cross-sectional dependence test results in Table B.2 indicate cross-sectional dependence in the real-wage data.[17] Therefore, a second-generation unit-root test is appropriate.

[17] The cross-sectional dependence tests are based on: (a) Pesaran, M. H. (2015). Testing Weak Cross-Sectional Dependence in Large Panels. *Econometric Reviews*, 34(6-10), 1089-1117; (b) Pesaran, M. H. (2021). General diagnostic tests for cross-sectional dependence in panels. *Empirical Economics*, 60: 13-50; and (c) Pesaran, M. H. & Xie, Y. (2021). A Bias-Corrected CD Test for Error Cross- Sectional Dependence in Panel Data Models with Latent Factors. Cambridge Working Papers in Economics 2158. The software module 'xtcd2' for Stata prepared by Jan Detizen has been used.

Accordingly, we tested the unit root of the real wage rate using Pesaran's CIPS (Cross-sectionally Augmented IPS) test.[18] For this test, we include two period lags; we include a trend when the test is run in levels. The test statistic in levels is below the critical value, and the test statistic in differences is above the critical limit. Thus, based on these results, real wages may be taken as I(1), which matches the trends in this variable shown in Table 2.

The results of the cross-sectional dependence test for RFWPR are mixed, but the variable may be marked by cross-sectional dependence. This also applies to RMPCE. Pesaran's CIPS test results for RFWPR indicate that the variable is I(1). The Hadri test indicates that RFWPR is I(1), and the CIPS test also supports this; therefore, the variable is I(1). Two first-generation tests for RMPCE in Table B.1 indicate that the variable is I(1). The CIPS test results also point in that direction. Thus, RMPCE may be taken as I(1).

[18] Pesaran, M.H. (2007). A simple panel unit root test in the presence of cross-section dependence. *Journal of Applied Econometrics*, 22(2), 265-312. The critical value shown in Table B.2 is an approximation, because our data has a much larger number of cross-sectional units than the number for which the critical values are given in Pesaran's paper.